\documentclass[aps,english,onecolumn,notitlepage]{revtex4-1}
\usepackage{epsfig}
\usepackage{amsmath}
\usepackage{amssymb}
\usepackage{framed}
\usepackage{mathtools}

\DeclarePairedDelimiterX\braket[2]{\langle}{\rangle}{#1 \delimsize\vert #2}
\usepackage{hyperref}
\usepackage[rightcaption]{sidecap}
\usepackage[figuresleft]{rotating}
\usepackage{caption}
\usepackage{float}
\usepackage{color,pstcol}
\usepackage{subfigure}
\usepackage{graphicx}

\begin{document}
\title{Spin-flip scattering engendered quantum spin torque in a Josephson junction} \author{Subhajit Pal} \author{Colin Benjamin} \email{colin.nano@gmail.com}\affiliation{School of Physical Sciences, National Institute of Science Education \& Research, HBNI, Jatni-752050,\ India }
\begin{abstract}
We examine a Josephson junction with two ferromagnet's and a magnetic impurity sandwiched between two superconductors. In such ferromagnetic Josephson junctions equilibrium spin torque exists only when ferromagnet's are misaligned. This is explained via the \textquotedblleft conventional\textquotedblright{} mechanism of spin transfer torque, which owes its origin to the misalignment of two ferromagnet's. However, we see surprisingly when the magnetic moments of the ferromagnet's are aligned parallel or anti-parallel, there is a finite equilibrium spin torque due to the quantum mechanism of spin flip scattering. We explore the properties of this unique spin flip scattering induced equilibrium quantum spin torque, especially its tunability via exchange coupling and phase difference across the superconductors. \end{abstract}\maketitle
\section{Introduction}
When a spin polarized current enters a ferromagnetic layer, there is generally a transfer of spin angular momentum between the conduction electrons and the magnetization of the ferromagnet. This was first proposed by Slonczewski\cite{slon} and Berger\cite{Ber} in 1996 as a novel mechanism for switching the magnetization of a ferromagnet by a spin polarized current. It was experimentally realized in spin-valve trilayers in 2000\cite{kat}. Since then, spin transfer torque has been investigated in various magnetic nanostructures\cite{Zha,Baek}. In a spin valve, when electric current passes through a fixed magnetic layer, it becomes spin polarized along the direction of the magnetic moment of the fixed magnetic layer. After passing through a nonmagnetic metal layer, the current enters into the free magnetic layer and polarizes along the magnetization direction of the free magnetic layer. When the magnetic moments of the two magnetic layers are not parallel or anti-parallel, free magnetic layer can absorb the spin polarized current\cite{Stil}. Due to this absorption, some angular momentum can be transferred to the free layer. Thus, a torque arises on the magnetic moment of the free layer which can cause the switching of the free layer's  magnetization. The aforesaid torque is generally described as non-equilibrium spin transfer torque since it needs a voltage bias to operationalize it. The spin transfer torque can also arise in equilibrium situation in absence of a voltage bias as in Josephson junction.\par
In ferromagnetic Josephson junction's\cite{buz,mey}, the Josephson super-current induces an equilibrium spin transfer torque due to the misaligned magnetic moments of the ferromagnetic layers\cite{Wain} which is proportional to the sine of the difference in magnetization directions of the two ferromagnet's. If $\mathcal{F}$ is the free energy of the superconductor-ferromagnet-normal metal-ferromagnet-superconductor ($SF_{1}NF_{2}S$) junction and $\theta$ is the angle between the magnetic moments of the ferromagnet's, then equilibrium spin transfer torque is defined as\cite{Wain}- $\tau^{eq}=\frac{\partial \mathcal{F}}{\partial \theta}$,
\begin{figure}[h]
\vskip -0.16 in
\centering{\includegraphics[width=.9\linewidth]{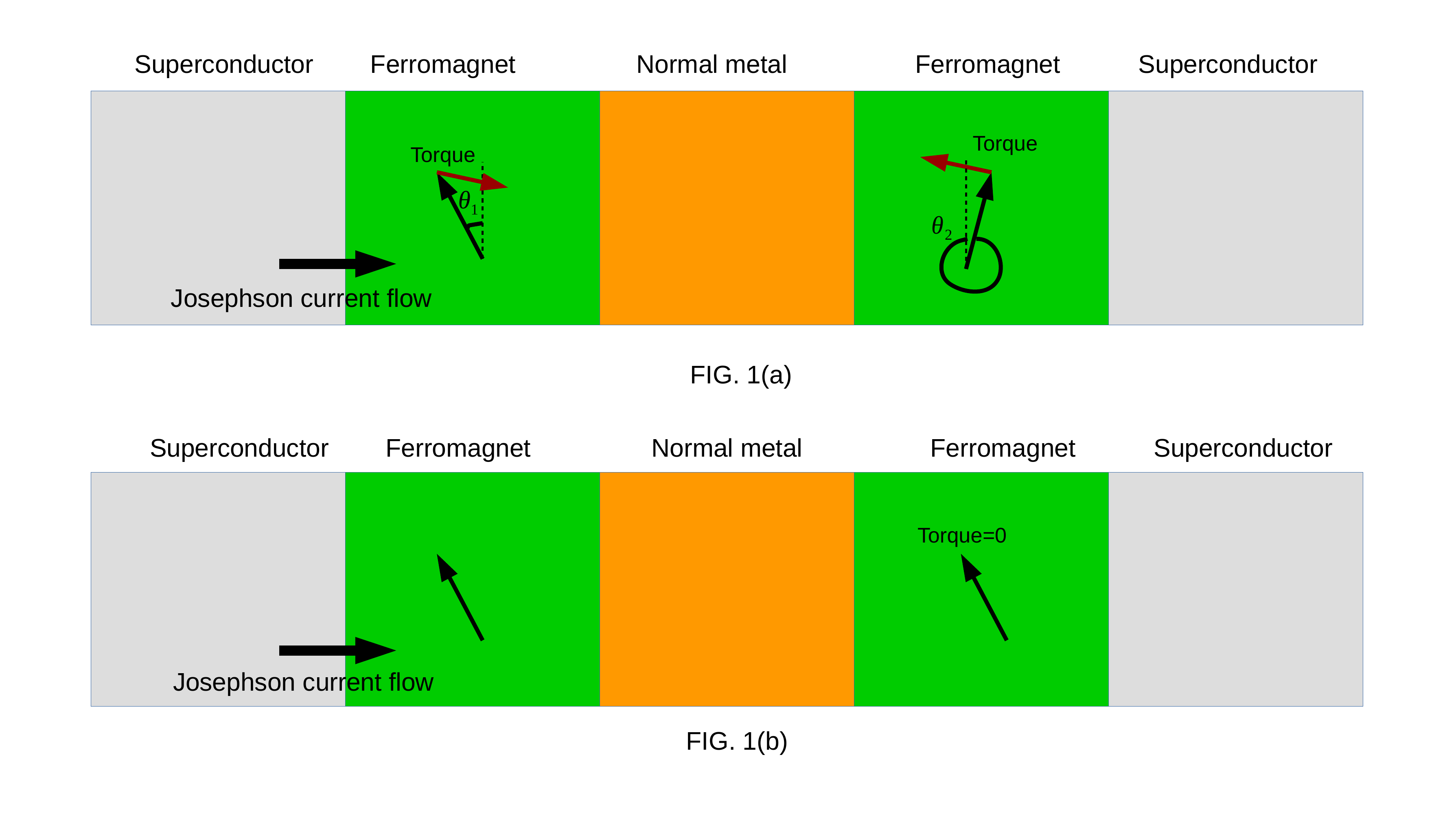}}
\caption{\small\sl Conventional mechanism of the equilibrium spin transfer torque in a superconductor-ferromagnet-normal metal-ferromagnet-superconductor junction. (a) Magnetic moments of the ferromagnet's are misaligned ($\theta_{1} \neq \theta_{2}$).  
Equilibrium spin transfer torque $\tau^{eq} \propto \sin (\theta_{1}-\theta_{2})$ and points perpendicular to the plane spanned by the two magnetic moments of the ferromagnet's, (b) Magnetic moments of the ferromagnet's are aligned ($\theta_1 = \theta_2  $). $\tau^{eq}=0$: equilibrium spin transfer torque vanishes.}
\end{figure}
with Josephson super-current\cite{deG}, $I=\frac{2e}{\hbar}\frac{\partial \mathcal{F}}{\partial \varphi}$, and $\varphi$ is the phase difference between the two superconductors. Thus,- $\frac{\partial I}{\partial \theta}=\frac{2e}{\hbar}\frac{\partial \tau^{eq}}{\partial \varphi}$, 
which relates Josephson current to equilibrium spin transfer torque. The Josephson super-current, similar to the diagram shown in Fig.~1, depends on sine of phase difference across superconductors $(\varphi_{L}-\varphi_{R})$ and flows from left to right or vice-versa. Equilibrium spin transfer torque points perpendicular to the plane spanned by the two magnetic moments of the ferromagnetic layers\cite{Wain} and its magnitude is sinusoidal in difference in magnetization directions of the two ferromagnet's. Sign and magnitude of the equilibrium spin transfer torque can be controlled by the phase difference between the two superconductors\cite{Wain}.\par
The equilibrium spin torque seen previously in SFFS junction\cite{Wain} or SFFFS junction\cite{Halter} or even SFSFS junction\cite{halter} is due to misalignment of ferromagnet's. The origin of equilibrium spin transfer torque is \textquotedblleft classical\textquotedblright. This can be easily understood via a \textquotedblleft classical mechanism\textquotedblright, see Fig.~1(a). But, this conventional view of the origin of spin transfer torque may not be always applicable. The quantum origins of spin torque, as opposed to the \textquotedblleft classical\textquotedblright spin transfer torque have been speculated recently in Refs.~\cite{spin1,spin2}. In this paper, we give an example where the mechanism underlying the equilibrium spin torque is quantum in nature and due to spin flip scattering. Classically, when the magnetic moment of electron is parallel or anti-parallel to the magnetic field, there is no torque exerted on the electron. Similarly, when the two magnetic moments of the ferromagnetic layers in a Josephson junction are parallel or anti-parallel equilibrium spin transfer torque vanishes, see Fig.~1(b). In Ref.~\cite{Wain}, the equilibrium spin transfer torque also follows the same behavior. But in this paper our main motivation is to show that if we replace the normal metal of Ref.\cite{Wain} by a magnetic impurity between two ferromagnetic layers, we will see a new effect- existence of a finite equilibrium spin torque even when magnetic moments of the ferromagnet's are aligned parallel or anti-parallel. We show that a magnetic impurity can engender a torque in such a junction. We call this \textquotedblleft equilibrium quantum spin torque\textquotedblright. Thus, this new mechanism of spin flip scattering can lead to finite equilibrium spin torque which has no classical analog.\par
The reason we are interested in spin transfer torque is because of the manifold applications like switching of the magnetization of ferromagnet's for sufficiently large current without any external magnetic field. This switching provides a mechanism to create fast magnetic random access memories\cite{Myer}. Further spin transfer torque can also be used for excitation of spin waves\cite{tsoi}. The equilibrium spin transfer torque first shown in Ref.~\cite{Wain} with $s$-wave superconductor has been extended to $d$-wave in Ref.~\cite{lin}.\par
The paper is organized as follows: in the next section on Theory, we first present our model and discuss the theoretical background of our study by writing the Hamiltonian, wave-functions and boundary conditions needed to calculate charge Josephson current and equilibrium quantum spin torque. In section III, we analyze our results for equilibrium quantum spin torque (subsection III.~A) and discuss the physical picture of torque (subsection III.~B). In section IV, we give an experimental realization and brief conclusion to our study. The explicit form of expression of equilibrium quantum spin torque is provided in the Appendix.         
\section{Theory}
The Hamiltonian, wave-functions, boundary conditions of our system as depicted in Fig.~2 and the calculations of Andreev bound states are done in this section.
\subsection{Hamiltonian}
Our system consists of two ferromagnet's ($F_{1}$ and $F_{2}$) with a {magnetic impurity}, sandwiched between two conventional s-wave singlet superconductors. The superconductors are isotropic and our model is shown in Fig.~2, with a {magnetic impurity} at $x=0$, two s-wave superconductors on either side at $x<-a/2$ and $x>a/2$ and two ferromagnetic layers in regions: $-a/2<x<0$ and $0<x<a/2$. In general, $h$ the magnetization vectors of the two ferromagnetic layers are misaligned by an angle $\theta$. However, in our calculation we focus on the limit $\theta\rightarrow0$, i.e., magnetization vectors are aligned parallely. We take the superconducting pair potential of the form $\Delta=\Delta_{0}[e^{i\varphi_{L}}\Theta(-x-a/2)+e^{i\varphi_{R}}\Theta(x-a/2)]$, where $\Delta_{0}$ is gap parameter, $\varphi_{L}$ and $\varphi_{R}$ are the superconducting phases for left and right superconductor respectively. The temperature dependence of $\Delta_{0}$ is given by $\Delta_{0}\rightarrow \Delta_{0}\tanh(1.74\sqrt{(T_{c}/T-1)})$, where $T_{c}$ is the superconducting critical temperature\cite{annu}.
\begin{figure}[h]\vskip -0.16in
\centering{\includegraphics[width=1.1\linewidth]{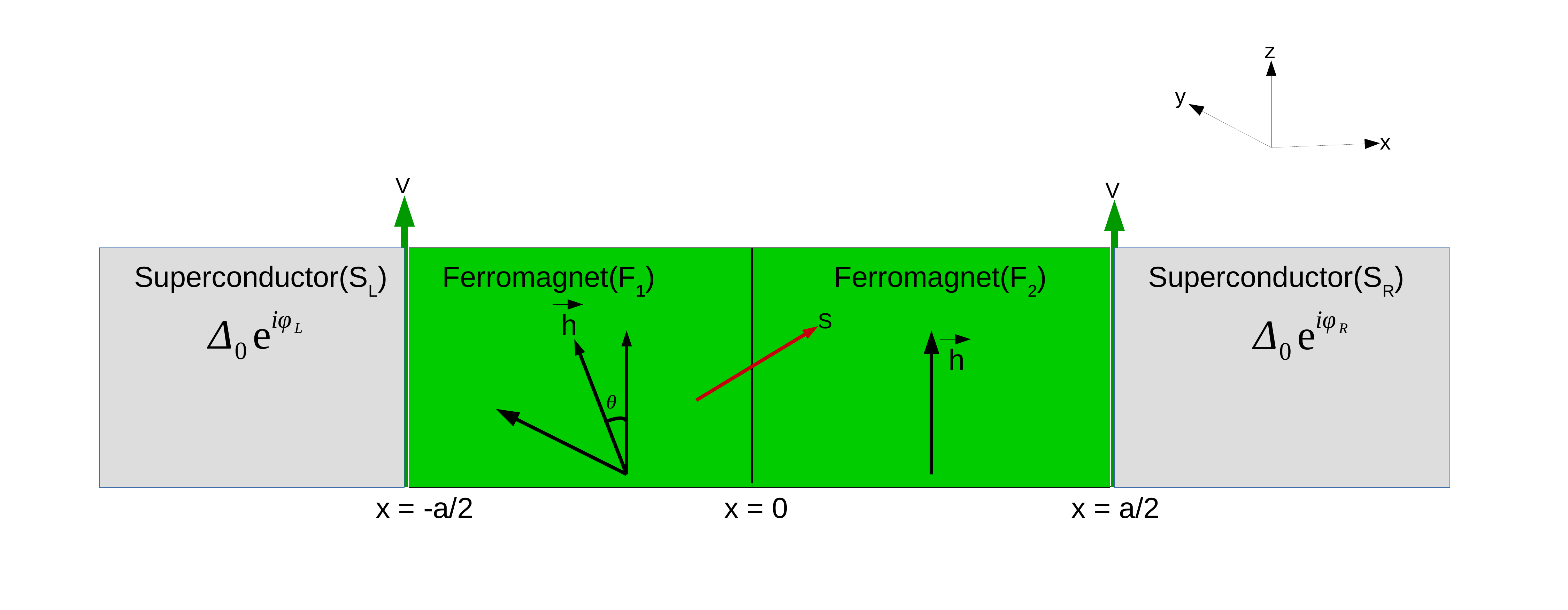}}
\caption{\small\sl Josephson junction composed of two ferromagnet's and a {magnetic impurity} with spin $S$ and magnetic moment $m'$ at $x=0$ sandwiched between two s-wave superconductors. In our work $\theta_{1}=\theta$ and $\theta_{2}=0$.  When ferromagnet's are aligned, i.e., $\theta\rightarrow0$, equilibrium spin transfer torque vanishes (see Fig.~1(b)), however in our setup a new quantum mechanism of spin flip scattering gives rise to a non-zero torque, which we denote as Equilibrium quantum spin torque (EQST). In this paper, we mainly concentrate on the limit $\theta\rightarrow0$.}
\end{figure}
The Bogoliubov-de Gennes equation of our system  is given below\cite{Liu,cheng}:
\begin{eqnarray}
\begin{pmatrix}
H_{0}\hat{I} & i\Delta \hat{\sigma}_{y}\\
-i\Delta^{*} \hat{\sigma}_{y} & -H_{0}^{*}\hat{I}
\end{pmatrix} \psi(x)& =& E \psi(x), 
\label{eqq}
\end{eqnarray}
where $H_{0}=p^2/2m^{\star}+V[\delta(x+a/2)+\delta(x-a/2)]-J_{0}\delta(x)\vec s.\vec S-\vec{h}.\hat{\sigma}[\Theta(x+a/2)+\Theta(a/2-x)]-E_{F}$. In the Hamiltonian \textquotedblleft$H_{0}$\textquotedblright{}, the first term describes the kinetic energy of an electron with mass $m^{\star}$, the second term depicts interfaces- $V$ is the strength of the $\delta$-like potential at the interfaces between ferromagnet and superconductor, the third term describes {magnetic impurity} with $J_{0}$ being the strength of exchange interaction between the electron with spin $\vec{s}$ and the {magnetic impurity} with spin $\vec{S}$\cite{joseph,AJP}, the fourth term describes ferromagnet's with $\vec{h}$ being the magnetization vectors of the two ferromagnet's and $\Theta$ is the Heaviside step function. Further, $\psi(x)$ is a four-component spinor, $E_{F}$ is the Fermi energy, $\hat{\sigma}$ is the Pauli spin matrix and $\hat{I}$ is the $2\times2$ identity matrix. In general, the magnetization vector ($\vec{h}$) of left ferromagnet ($F_{1}$) is assumed to be at an angle of $\theta$ with $z$ axis in the $y-z$ plane, while that of right ferromagnet ($F_{2}$) is fixed along the $z$ axis. Thus, $\vec{h}.\hat{\sigma}=h\sin \theta\hat{\sigma}_{y}+h\cos \theta\hat{\sigma}_{z}$\cite{Halter}. However, in our study we only concentrate on the case where $\theta\rightarrow 0$, i.e., Ferromagnet's are aligned. In the subsequent analysis we take the dimensionless version of $J_{0}$ and $V$ given as $J=\frac{m^{*}J_{0}}{\hbar^2k_{F}}$ and $Z=\frac{m^{*}V}{\hbar^2k_{F}}$\cite{BTK}. 
\subsection{Wave-functions and boundary conditions in the ferromagnetic Josephson junction in presence of a {magnetic impurity}}
The system we consider consists of two ferromagnet's with a {magnetic impurity} sandwiched between two conventional s-wave singlet superconductors.
Our model is shown in Fig.~2, it depicts a {magnetic impurity} at $x=0$ and two superconductors at $x<-a/2$ and $x>a/2$. There are two ferromagnetic regions in $-a/2<x<0$ and $0<x<a/2$.\par
\subsubsection{Wave-functions}
Let us consider a spin up electron incident at $x=-a/2$ interface from left superconductor. 
If we solve the Bogoliubov-de Gennes equation  for superconductors (see Eq.~\ref{eqq}), we will get the wavefunctions for left and right superconductors. 
The wave function in the left superconductor (for $x<-\frac{a}{2}$) is \cite{joseph,LINDER}-
\begin{eqnarray}
\label{eq1}
\psi_{S_{L}}(x)=\begin{pmatrix}u\\
                              0\\
                              0\\ 
                              v
                              \end{pmatrix}e^{ik_{+}x}\phi_{m'}^{S}+r_{ee}^{\uparrow\uparrow}\begin{pmatrix}
                              u\\
                              0\\
                              0\\
                              v
                             \end{pmatrix}e^{-ik_{+}x}\phi_{m'}^{S}+r_{ee}^{\uparrow\downarrow}\begin{pmatrix}
                             0\\
                             u\\
                            -v\\
                             0
                             \end{pmatrix}e^{-ik_{+}x}\phi_{m'+1}^{S}+r_{eh}^{\uparrow\uparrow}\begin{pmatrix}
                             0\\
                            -v\\
                             u\\
                             0
                             \end{pmatrix}e^{ik_{-}x}\phi_{m'+1}^{S}+r_{eh}^{\uparrow\downarrow}\begin{pmatrix}
                             v\\
                             0\\
                             0\\
                             u
                             \end{pmatrix}e^{ik_{-}x}\phi_{m'}^{S},\nonumber\\
                             \end{eqnarray}
The amplitudes $r_{ee}^{\uparrow\uparrow},r_{ee}^{\uparrow\downarrow},r_{eh}^{\uparrow\uparrow},r_{eh}^{\uparrow\downarrow}$ are normal reflection without flip, normal reflection with spin flip, Andreev reflection with spin flip and Andreev reflection without flip respectively. The corresponding wave function in the right superconductor (for $x>\frac{a}{2}$) is given by-
\begin{eqnarray}
\psi_{S_{R}}(x)=t_{ee}^{\uparrow\uparrow}\begin{pmatrix}
                              ue^{i\varphi}\\
                              0\\
                              0\\
                              v
                             \end{pmatrix}e^{ik_{+}x}\phi_{m'}^{S}+t_{ee}^{\uparrow\downarrow}\begin{pmatrix}
                             0\\
                             ue^{i\varphi}\\
                             -v\\
                             0
                             \end{pmatrix}e^{ik_{+}x}\phi_{m'+1}^{S}+t_{eh}^{\uparrow\uparrow}\begin{pmatrix}
                             0\\
                             -ve^{i\varphi}\\
                             u\\
                             0
                             \end{pmatrix}e^{-ik_{-}x}\phi_{m'+1}^{S}+t_{eh}^{\uparrow\downarrow}\begin{pmatrix}
                             ve^{i\varphi}\\
                             0\\
                             0\\
                             u
                             \end{pmatrix}e^{-ik_{-}x}\phi_{m'}^{S},\nonumber\\
                             \end{eqnarray}
where $t_{ee}^{\uparrow\uparrow},t_{ee}^{\uparrow\downarrow},t_{eh}^{\uparrow\uparrow},t_{eh}^{\uparrow\downarrow}$ represent transmission amplitudes, corresponding to the reflection process described above and $\varphi=\varphi_{R}-\varphi_{L}$ represents the phase difference between right and left superconductors. $\phi_{m'}^{S}$ is the eigenspinor of the {magnetic impurity}, with its $S^{z}$ operator acting as- $S^{z}\phi_{m'}^{S} = m'\phi_{m'}^{S}$, with $m'$ being the spin magnetic moment of the {magnetic impurity}. $u$ and $v$ are the BCS coherence factors which are defined as $u^{2}={\frac{1}{2}(1+\frac{\sqrt{E^2-\Delta_{0}^{2}}}{E})}$, $v^{2}={\frac{1}{2}(1-\frac{\sqrt{E^2-\Delta_{0}^{2}}}{E})}$.
$k_{\pm}=\sqrt{\frac{2m^{\star}}{\hbar^2}(E_{F}\pm \sqrt{E^2-\Delta_{0}^2})}$ is the wave-vector for electron-like quasi-particle ($k_{+}$) and hole-like quasi-particle ($k_{-}$) in the left and right superconducting wave-functions, $\psi_{S_L}$ and $\psi_{S_R}$.\\
Similarly solving the Bogoliubov-de Gennes equation for ferromagnet's, we get the wave-function in ferromagnet's.
The wave-function in the left ferromagnet ($F_{1}$) is given by-
\begin{eqnarray}
\psi_{F_{1}}(x)=(ee^{iq_{\uparrow}^{+}(x+a/2)}+fe^{-iq_{\uparrow}^{+}x})\begin{pmatrix}
                                                     \cos \frac{\theta}{2}\\
                                                     i\sin\frac{\theta}{2}\\
                                                     0\\
                                                     0
                                                    \end{pmatrix}\phi_{m'}^{S}+(e^{\prime} e^{iq_{\downarrow}^{+}(x+a/2)}+f^{\prime} e^{-iq_{\downarrow}^{+}x})\begin{pmatrix}
                                                     i\sin \frac{\theta}{2}\\
                                                     \cos\frac{\theta}{2}\\
                                                     0\\
                                                     0
                                                     \end{pmatrix}\phi_{m'+1}^{S}\nonumber\\+(e_{0}e^{-iq_{\uparrow}^{-}(x+a/2)}+f_{0}e^{iq_{\uparrow}^{-}x})\begin{pmatrix}
                                                     0\\
                                                     0\\
                                                     \cos\frac{\theta}{2}\\
                                                     -i\sin\frac{\theta}{2}
                                                     \end{pmatrix}\phi_{m'+1}^{S}+(e_{0}^{\prime} e^{-iq_{\downarrow}^{-}(x+a/2)}+f_{0}^{\prime} e^{iq_{\downarrow}^{-}x})\begin{pmatrix}
                                                     0\\
                                                     0\\
                                                     -i\sin\frac{\theta}{2}\\
                                                     \cos\frac{\theta}{2}
                                                     \end{pmatrix}\phi_{m'}^{S},\mbox{for $-\frac{a}{2}<x<0.$}
                                                     \end{eqnarray}
Similarly, the wave-function in the right ferromagnet ($F_{2}$) is given by-
\begin{eqnarray}
\psi_{F_{2}}(x)=(a_{0}e^{iq_{\uparrow}^{+}x}+be^{-iq_{\uparrow}^{+}(x-a/2)})\begin{pmatrix}
                                                     1\\
                                                     0\\
                                                     0\\
                                                     0
                                                    \end{pmatrix}\phi_{m'}^{S}+(a^{\prime} e^{iq_{\downarrow}^{+}x}+b^{\prime} e^{-iq_{\downarrow}^{+}(x-a/2)})\begin{pmatrix}
                                                     0\\
                                                     1\\
                                                     0\\
                                                     0
                                                     \end{pmatrix}\phi_{m'+1}^{S}\nonumber\\+(ce^{-iq_{\uparrow}^{-}x}+de^{iq_{\uparrow}^{-}(x-a/2)})\begin{pmatrix}
                                                     0\\
                                                     0\\
                                                     1\\
                                                     0
                                                     \end{pmatrix}\phi_{m'+1}^{S}+(c^{\prime} e^{-iq_{\downarrow}^{-}x}+d^{\prime} e^{iq_{\downarrow}^{-}(x-a/2)})\begin{pmatrix}
                                                     0\\
                                                     0\\
                                                     0\\
                                                     1
                                                     \end{pmatrix}\phi_{m'}^{S},\mbox{for $0<x<\frac{a}{2}.$}
                                                    \end{eqnarray}    
$q_{\sigma}^{\pm}=\sqrt{\frac{2m^{\star}}{\hbar^2}(E_{F}+\rho_{\sigma}h\pm E)}$ is the wave-vector for electron ($q_{\sigma}^{+}$) and hole ($q_{\sigma}^{-}$) in the ferromagnetic layers, wherein $\rho_{\sigma}=+1(-1)$ is related to $\sigma=\uparrow(\downarrow)$. In our work we have used the Andreev approximation $k_{+}=k_{-}=\sqrt{\frac{2m^{\star}E_{F}}{\hbar^2}}=k_{F}$ and $q_{\uparrow,\downarrow}=k_{F}\sqrt{1\pm\frac{h}{E_{F}}}$, where $k_{F}$ is the Fermi wave-vector, with $E_{F}>>\Delta, E$.\par
\subsubsection{Boundary conditions}
The boundary conditions can be written as follows\cite{joseph,LINDER}: at $x=-a/2$- $\psi_{S_{L}}(x)=\psi_{F_{1}}(x)$ (continuity of wave-functions) and $\frac{d\psi_{F_{1}}}{dx}-\frac{d\psi_{S_{L}}}{dx}=\frac{2m^{\star}V}{\hbar^2}\psi_{F_{1}}$, (discontinuity in first derivative), at $x=0$ (see Fig.~1), $\psi_{F_{1}}(x)=\psi_{F_{2}}(x)$ and $\frac{d\psi_{F_{2}}}{dx}-\frac{d\psi_{F_{1}}}{dx}=-\frac{2m^{\star}J_{0}\vec s.\vec S}{\hbar^2} \psi_{F_{1}}$ with $\vec s.\vec S=s^{z}S^{z}+\frac{1}{2}(s^{-}S^{+}+s^{+}S^{-})$ being the exchange coupling due to {magnetic impurity} in the Hamiltonian. $\vec s$ represents spin operator acting on electron/hole states $\phi_{m}^{s}$, while $\vec S$ represents the spin operator acting on {magnetic impurity} states $\phi_{m'}^{S}$. $\phi_{m}^{s}$ and $\phi_{m'}^{S}$ are the eigenstates of electron/hole and magnetic impurity with $m$ and $m'$ being the spin magnetic moment of the electron/hole and {magnetic impurity} respectively. $s$ is spin of electron, while $S$ is spin of {magnetic impurity}, with $s^{\pm}=s^{x}\pm i s^{y}, s^{z}=\frac{\hbar}{2}\begin{pmatrix}\sigma_{z} & 0 \\
0 & \sigma_{z} \end{pmatrix}, s^{x}=\frac{\hbar}{2}\begin{pmatrix}  \sigma_{x} & 0\\
0 & \sigma_{x} \end{pmatrix}, s^{y}=\frac{\hbar}{2}\begin{pmatrix}  \sigma_{y} & 0\\
0 & \sigma_{y} \end{pmatrix} $, where $\sigma_{z}=\begin{pmatrix}1 & 0 \\
0 & -1 \end{pmatrix}, \sigma_{x}=\begin{pmatrix}0 & 1 \\
1 & 0 \end{pmatrix}, \sigma_{y}=\begin{pmatrix}0 & -i \\
i & 0 \end{pmatrix}$. The action of spin raising and spin lowering operators for {magnetic impurity} are discussed below.
For spin up electron incident $\phi_{m}^{s}=(1\hspace{2pt}0\hspace{2pt}0\hspace{2pt}0)^{T}   $, with $s=1/2$, $m=1/2$, T denotes transpose of matrix.
Now, when the spin flip term of our Hamiltonian acts on spin up electron $(1\hspace{2pt}0\hspace{2pt}0\hspace{2pt}0)^{T}$, where $T$ stands for transpose and the {magnetic impurity} state $\phi_{m'}^{S}$ we have-
\begin{equation}
\vec{s}.\vec{S} (1\hspace{2pt}0\hspace{2pt}0\hspace{2pt}0)^{T}\phi_{m'}^{S}=s^{z}S^{z}(1\hspace{2pt}0\hspace{2pt}0\hspace{2pt}0)^{T}\phi_{m'}^{S}+\frac{1}{2}s^{-}S^{+}(1\hspace{2pt}0\hspace{2pt}0\hspace{2pt}0)^{T}\phi_{m'}^{S}+\frac{1}{2}s^{+}S^{-}(1\hspace{2pt}0\hspace{2pt}0\hspace{2pt}0)^{T}\phi_{m'}^{S}
\label{eqq1}
\end{equation}
Now, $s^{+}(1\hspace{2pt}0\hspace{2pt}0\hspace{2pt}0)^{T}=0$, since $s^{+}$ is the spin raising operator and there are no higher spin states for a spin-$1/2$ electron than up and so the 3rd term in Eq.~\ref{eqq1} vanishes, while $s^{-}(1\hspace{2pt}0\hspace{2pt}0\hspace{2pt}0)^{T}=(0\hspace{2pt}1\hspace{2pt}0\hspace{2pt}0)^{T}$, the spin lowering operator gives the down spin state $(0\hspace{2pt}1\hspace{2pt}0\hspace{2pt}0)^{T}$ of electron.  Further, for spin-up electron $s^{z} (1\hspace{2pt}0\hspace{2pt}0\hspace{2pt}0)^{T}= \frac{1}{2} (1\hspace{2pt}0\hspace{2pt}0\hspace{2pt}0)^{T}$, as $\hbar=1$ and for {magnetic impurity}- $S^{z} \phi_{m'}^{S}=m' \phi_{m'}^{S}$. Further, the spin-raising and spin-lowering operators acting on {magnetic impurity} give:$S^{+}\phi_{m'}^{S}=f_{2}\phi_{m'+1}^S=\sqrt{(S-m')(S+m'+1)}\phi_{m'+1}^S$ and $S^{-}\phi_{m'+1}^{S}=\sqrt{(S-m')(S+m'+1)}\phi_{m'}^S$.
\begin{equation}
\mbox{ Thus, }\vec{s}.\vec{S}(1\hspace{2pt}0\hspace{2pt}0\hspace{2pt}0)^{T}\phi_{m'}^{S}= \frac{1}{2} m'(1\hspace{2pt}0\hspace{2pt}0\hspace{2pt}0)^{T}\phi_{m'}^{S}+\frac{1}{2}\sqrt{(S-m')(S+m'+1)}(0\hspace{2pt}1\hspace{2pt}0\hspace{2pt}0)^{T}\phi_{m'+1}^{S}
\label{eqq2}
\end{equation}
From Eqs.~\ref{eqq1}, \ref{eqq2} we thus have-
\begin{equation}
\vec{s}.\vec{S}(1\hspace{2pt}0\hspace{2pt}0\hspace{2pt}0)^{T}\phi_{m'}^{S}=\frac{1}{2}m'(1\hspace{2pt}0\hspace{2pt}0\hspace{2pt}0)^{T}\phi_{m'}^{S}+\frac{1}{2}f_{2}(0\hspace{2pt}1\hspace{2pt}0\hspace{2pt}0)^{T}\phi_{m'+1}^{S}{} \mbox{ (for both no flip and spin flip process) }\nonumber 
\end{equation}
In quantum spin flip scattering process when { spin polarized super-current (the state of spin polarized super-current is given as-$|s.c\rangle$), in our case denoted by a macroscopic wave-function $\sim|\Psi_{S_{K}}|e^{i\varphi_{K}}\approx \begin{pmatrix}u\\
                              0\\
                              0\\ 
                              v
                              \end{pmatrix}e^{i\varphi_{K}}$ (where $K$ can be $L$ or $R$)}, interacts with the {magnetic impurity}, the {magnetic impurity} can flip its spin with finite probability, but there is no certainty for flipping its spin. In addition to the spin flip process, there can be the other process without any flip. Thus, { the spin polarized super-current-{magnetic impurity} state} after exchange interaction is in a superposition of mutual spin-flip as well as no flip state given by the joint entangled wave-function of { spin polarized super-current (s.c)} and {magnetic impurity} as-
\begin{equation}
|s.c\rangle\otimes|\phi_{m'}^{S}\rangle={\frac{m'}{2}}|{\mbox{No flip}}\rangle+{\frac{f_{2}}{2}}|{\mbox{Mutual-flip}}\rangle\nonumber
\end{equation}
On the other hand when there is no possibility of spin-flip scattering, i.e., when $S=m'$, then spin flip probability of: $f_{2}$=$\sqrt{(S-m')(S+m'+1)}$=$0$ and $\vec{s}.\vec{S}(1\hspace{2pt}0\hspace{2pt}0\hspace{2pt}0)^{T}\phi_{m'}^{S}=s^{z}S^{z}\phi_{m'}^{S}= \frac{1}{2} m'(1\hspace{2pt}0\hspace{2pt}0\hspace{2pt}0)^{T}\phi_{m'}^{S}.$ Thus both before as well as after the spin polarized super-current state and magnetic impurity state are not entangled and neither there is any superposition.
{Thus, Hamiltonian $H_{0}=p^2/2m^{\star}+V[\delta(x+a/2)+\delta(x-a/2)]-J_{0}\delta(x)s^{z}S^{z}-\vec{h}.\hat{\sigma}[\Theta(x+a/2)+\Theta(a/2-x)]-E_{F}$, for only no flip process, while it is, $H_{0}=p^2/2m^{\star}+V[\delta(x+a/2)+\delta(x-a/2)]-J_{0}\delta(x)\vec s.\vec S-\vec{h}.\hat{\sigma}[\Theta(x+a/2)+\Theta(a/2-x)]-E_{F}$, for the case wherein mutual spin flip takes place with finite probability.}
Finally, at $x=a/2$, the boundary conditions are- $\psi_{F_{2}}(x)=\psi_{S_{R}}(x)$, $\frac{d\psi_{S_{R}}}{dx}-\frac{d\psi_{F_{2}}}{dx}=\frac{2m^{\star}V}{\hbar^2}\psi_{F_{2}}$.\par
The aforesaid method of addressing spin-flip scattering process is not unique to our work, { many other papers have used the same model of spin flip scattering in different context}, mention may be made of- the first paper which introduced this model, see Ref.~\cite{AJP}, to model of quantum spin flip scattering in graphene, see Ref.~\cite{Maruri}, in modeling the quantum spin flip scattering  in a Josephson junction, see Ref.~\cite{joseph},
and finally in  modeling the occurrence of {Yu-Shiba-Rusinov (YSR)} bound states in  at the interface of normal metal-superconductor junction, see Ref.~\cite{ysr-pal}. 
\subsection{Andreev Bound states}
Following the procedure enunciated in Ref.~\cite{annu} to calculate bound state contribution to Josephson supercurrent we neglect the contribution from incoming quasiparticle, i.e., first term $\begin{pmatrix}u & 0 & 0 & v\end{pmatrix}^{T}e^{ik_{+}x}\phi_{m'}^{S}$ of Eq.~\ref{eq1}  and insert the wave functions in the boundary conditions, we get a homogeneous system of linear equations for the scattering amplitudes, $Qx=0$, where $x$ is a $8\times1$ column matrix and is given by $x=[r_{ee}^{\uparrow\uparrow},r_{ee}^{\uparrow\downarrow},r_{eh}^{\uparrow\uparrow},r_{eh}^{\uparrow\downarrow},t_{ee}^{\uparrow\uparrow},t_{ee}^{\uparrow\downarrow},t_{eh}^{\uparrow\uparrow},t_{eh}^{\uparrow\downarrow}]$ and $Q$ is a $8\times 8$ matrix obtained by expressing the scattering amplitudes in the two ferromagnetic layers by the scattering amplitudes in the left and right superconductor. For a nontrivial solution of this system, Determinant of $Q=0$. We thus get the Andreev bound state energy spectrum $E_{i}$, $i=\{1,...,8\}$\cite{Been}. This is the usual procedure for calculating the bound state spectra in Josephson junctions, see Refs.~\cite{annu}, \cite{LINDER}.
{We find that $E_{i}(i=1,...,8)=\pm\varepsilon_{p}(p=1,...,4)$.}
\subsection{Josephson charge current}
 On solving the boundary conditions, we have 8 Andreev bound states given as $E_{i}(i=1,...,8)=\pm\varepsilon_{p}(p=1,...,4)$.
From Andreev bound states energies\cite{Been}  we get the Free energy of our system, which is given by\cite{annu}:
{
\begin{eqnarray}
{}&&\mathcal{F}=-\frac{1}{\beta}\frac{1}{2\pi}\int_{0}^{2\pi}\ln\Big[\prod_{i}(1+e^{-\beta E_{i}})\Big]d(k_{F}a)\nonumber\\
  &&=-\frac{2}{\beta}\frac{1}{2\pi}\int_{0}^{2\pi}\sum_{p=1}^{4}\ln\Big[2\cosh\Big(\frac{\beta \varepsilon_{p}}{2}\Big)\Big]d(k_{F}a)
\label{ff}
\end{eqnarray}
}
{We consider only the short junction limit, i.e., $a<<\xi$, where $\xi$ is the superconducting coherence length and $a$ the width of the intervening ferromagnetic layers between superconductors, such that the total Josephson current is determined by only the bound state contribution, the continuum contribution is negligible and so neglected. See Refs.~[10,16] where similar to us the continuum contribution to the total Josephson current is also neglected in the short junction limit.}
The charge Josephson current at finite temperature is the derivative of the Free energy $\mathcal{F}$ of our system with respect to the phase difference $\varphi$ between left and right superconductors\cite{deG,LINDER},
\begin{equation}
I_{c}=\frac{2e}{\hbar}\frac{\partial \mathcal{F}}{\partial \varphi}=-\frac{2e}{\hbar}\frac{1}{2\pi}\int_{0}^{2\pi}\sum_{p=1}^{4}\tanh\Big(\frac{\beta \varepsilon_{p}}{2}\Big)\frac{\partial \varepsilon_{p}}{\partial \varphi}d(k_{F}a)
\label{ff1}
\end{equation}
herein $e$ is the electronic charge and $k_{F}a$ is the phase accumulated in ferromagnetic layers.\par
\subsection{Equilibrium spin torque}
From the Free energy of our system (Eq.~(\ref{ff})) we calculate the equilibrium spin torque\cite{Wain} by taking the derivative of the Free energy with respect to the misorientation angle \textquoteleft$\theta$\textquoteright (the angle between magnetic moments of the two ferromagnets)-
\begin{equation}
\tau^{eq}=\frac{\partial\mathcal{F}}{\partial \theta}=-\frac{1}{2\pi}\int_{0}^{2\pi}\sum_{p=1}^{4}\tanh\Big(\frac{\beta \varepsilon_{p}}{2}\Big)\frac{\partial \varepsilon_{p}}{\partial \theta}d(k_{F}a)
\label{ff2}
\end{equation}
Eqs.~\ref{ff1}, \ref{ff2} are the main working formulas of our paper. The equilibrium spin torque is also referred to as equilibrium spin current in some papers\cite{linder,halter}. In our calculation as previously mentioned we focus on the case where magnetization in two ferromagnet's is aligned, i.e., $\theta\rightarrow0$. In this limit we surprisingly see a finite equilibrium spin torque due to spin flip scattering upending, the classical reason behind spin torque being due to nonaligned magnetization. For transparent regime ($Z=0$) we find-
\begin{equation}
\tau^{eq}\mid_{\theta\rightarrow 0}=\frac{\Delta_{0}^2}{2\pi}\int_{0}^{2\pi}\Big(\tanh\Big(\frac{\beta M_{1}}{2}\Big){M_{1}^{\prime}}+\tanh\Big(\frac{\beta M_{2}}{2}\Big){M_{2}^{\prime}}+\tanh\Big(\frac{\beta M_{3}}{2}\Big){M_{3}^{\prime}}+\tanh\Big(\frac{\beta M_{4}}{2}\Big){M_{4}^{\prime}}\Big)d(k_{F}a)
\end{equation}
{where $M_{1}$, $M_{2}$, $M_{3}$, $M_{4}$, $M_{1}^{\prime}$, $M_{2}^{\prime}$, $M_{3}^{\prime}$ and $M_{4}^{\prime}$ are large expressions that depend on  exchange interaction ($J$), magnetization of the ferromagnet's, spin ($S$) and magnetic moment ($m'$) of {magnetic impurity}, the phase ($k_{F}a$) accumulated in ferromagnetic region and spin flip probability of {magnetic impurity} ($f_{2}$). Their explicit forms are given in Appendix. In Appendix we show that for no flip case, the EQST ($\tau^{eq}\mid_{\theta\rightarrow 0}$) vanishes. { In the next section from figures we will see that the EQST is zero in the limit $J\rightarrow0$ and $Z\rightarrow\infty$.}
\section{Results}
\subsection{Analysing EQST}
In Figs.~3-7, we analyze via various plots this unique quantum spin torque due to spin flip scattering alone. In Fig.~3 we plot both Josephson charge super-current as well as the equilibrium quantum spin torque (EQST) for different interface transparencies $Z$ as a function of the phase difference $\varphi$. 
\begin{figure}
\includegraphics[width=0.9\linewidth]{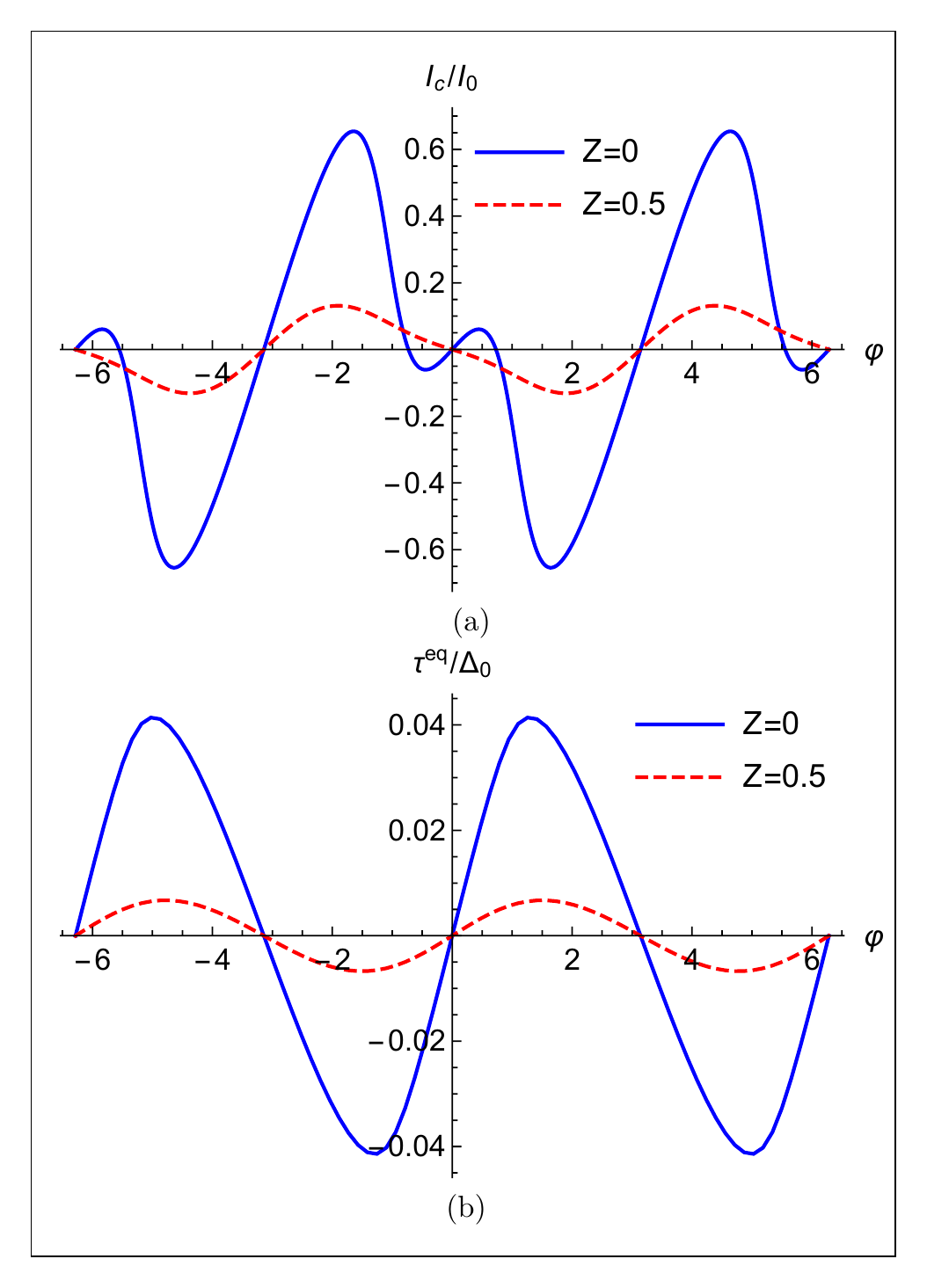}
\caption{\small \sl Josephson charge current and equilibrium quantum spin torque (EQST) as a function of phase difference ($\varphi$) for different values of interface barrier strength ($Z$). Parameters are $\Delta_{0}=1 meV$, $I_{0}=e\Delta_{0}/\hbar$, $T/T_{c}=0.01$, $J=0.5$, $h/E_{F}=0.5$, $\theta\rightarrow0$, $S=5/2$, $m'=-1/2$. { Both Josephson charge current and EQST are inhibited by increasing $Z$ while EQST is zero for $\varphi=0$ and $\varphi=2\pi$.}}
\end{figure}
We consider the magnetic moments of the ferromagnetic layers to be parallel ($\theta\rightarrow0$) and deal with the spin flip case, i.e., $f_{2}\neq 0$ (see Appendix), in this case $S\neq m'$ for {magnetic impurity} thus there is finite probability for {magnetic impurity} to flip its own spin while interacting with an electron/hole. We see both Josephson charge current and the EQST are inhibited by increasing interface barrier strength ($Z$). Further, similar to charge Josephson current, the EQST vanishes at $\varphi=0$ and $\varphi=2\pi$. Usually the spin transfer torque opposes the Josephson current (see Ref.~\cite{Wain}), however the equilibrium quantum spin torque (EQST) as shown here can flow in same direction as the Josephson current, see Fig.~3(a), $-0.7<\varphi<0.7$. This behavior is also seen in Ref.~\cite{spin2} for the quantum spin transfer torque in a different context.\par 
In Fig.~4(a) we plot the EQST as a function of phase difference ($\varphi$) for different values of exchange interaction $J$ again for $\theta\rightarrow0$. We see that with change of exchange interaction $J$ there is a sign change of EQST.  The change in sign of $\tau^{eq}$ via \textquoteleft$J$\textquoteright{} implies that the EQST seen in our system can be tuned via \textquoteleft$J$\textquoteright and the sign of $\tau^{eq}$ can be controlled by the phase difference as shown in Figs.~3(b), 4(a) \& 4(b). In Fig.~4(b) we plot EQST as function of phase difference ($\varphi$) for different values of the magnitude of magnetization ($h$) of the ferromagnet's. { We see that the EQST increases with increasing \textquoteleft$h$\textquoteright.}
\begin{figure}
\includegraphics[width=0.9\linewidth]{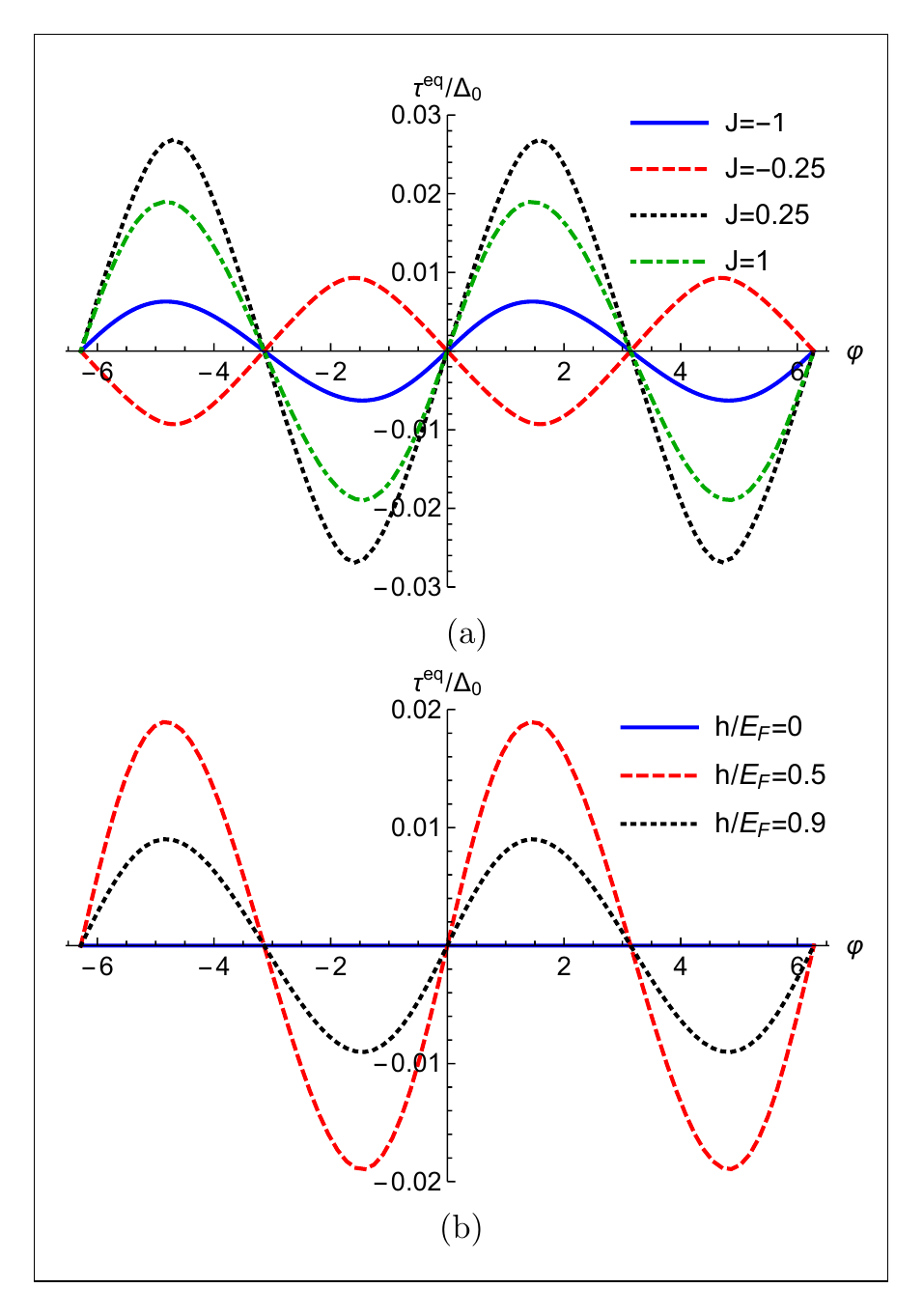}
\caption{\small \sl \small \sl EQST as a function of phase difference ($\varphi$) for (a) different values of exchange interaction ($J$) of {magnetic impurity} and for (b) different values of magnetization ($h$) of the ferromagnet's. Parameters are $\Delta_{0}=1 meV$, $I_{0}=e\Delta_{0}/\hbar$, $T/T_{c}=0.01$, $Z=0$, $J=1$ (for (b)), $h/E_{F}=0.5$, $\theta\rightarrow0$, $S=5/2$, $m'=-1/2$. { In (a) EQST changes sign with change of exchange interaction $J$ and phase difference $\varphi$. In (b) EQST increases with increasing magnetization $h$ of the ferromagnet's.}}
\end{figure}
In Fig.~5 we study the EQST from low to high spin states { and for different values of spin flip probability} of {magnetic impurity} again at $\theta\rightarrow0$ for a transparent junction, i.e., $Z=0$. In Fig.~5(a), $J=1$ and we see that the EQST monotonically decreases with increasing \textquoteleft$S$\textquoteright{} for particular value of $m'=-\frac{1}{2}$, implying high spin states inhibit EQST. In Fig. 5(b) we plot the EQST for a particular spin $S =5/2$ and for all possible values of spin flip probability of spin flipper. We see that EQST is enhanced for $f_{2}>S$ but for $f_{2}<S$ EQST is suppressed.\par
In Fig.~6(a) we plot the EQST for flip ($S=3/2, m'=-1/2, f_{2}\neq0$) case as well as no flip ($S=3/2, m'=3/2, f_{2}=0$) case and also for a superconductor-ferromagnet-ferromagnet-superconductor ($S$-$F_{1}$-$F_{2}$-$S$) junction without {magnetic impurity} ($J=0$) in the same figure as a function of mis-orientation angle ($\theta$) between ferromagnet's. We see that in contrast to $S$-$F_{1}$-$F_{2}$-$S$ junction ($J=0$ case) and no flip case, EQST is finite at $\theta\rightarrow0$ and $\theta=\pi$ when magnetic impurity flips its spin. Thus the reason for finite EQST at $\theta\rightarrow0$ is finite probability for flipping.  This can be explained as follows- after passing through first ferromagnetic layer the {super-current} become polarized in the direction of magnetization of the first ferromagnetic layer. When { spin polarized super-current} interacts with the {magnetic impurity} through the exchange interaction, there is a finite probability for a mutual spin flip.
The equation below depicts the interaction process:
\begin{equation}
|s.c\rangle\otimes|\phi_{m'}^{S}\rangle={\frac{m'}{2}}|{\mbox{No flip}}\rangle+{\frac{f_{2}}{2}}|{\mbox{Mutual-flip}}\rangle
\end{equation}
{where $|s.c\rangle$ is the state of spin polarized supercurrent}, see paragraphs above and below Eq.~\ref{eqq2} on how this aforesaid equation comes into being. Due to this spin flip scattering the direction of { the spin of supercurrent} will be in a superposition too and thus will differ from the direction of the magnetization vector of the ferromagnetic layer. Thus, when the {supercurrent} enters the second ferromagnetic layer, magnetization vector of the second ferromagnetic layer will exert a torque on the spin flipped component of the {supercurrent wave function} in order to rotate { the supercurrent's spin} along the direction of magnetization, while leaving the non spin flipped component as it is. From conservation of spin angular momentum, the {supercurrent} will also exert an equal and opposite torque on the magnetic moment of the second ferromagnetic layer leading to a finite EQST even at $\theta\rightarrow 0$.  However, in absence of {magnetic impurity} ($J=0$ case) and for no flip case the {spin polarized supercurrent state} does not flip it's spin. Thus, in absence magnetic impurity or in case of no-flip scattering the { spin polarized supercurrent's spin} and the magnetization vector of the ferromagnetic layers will be in the same direction. Therefore, EQST vanishes in case of $J=0$ and no-flip process but for spin-flip process it is finite.
\begin{figure}
\includegraphics[width=0.9\linewidth]{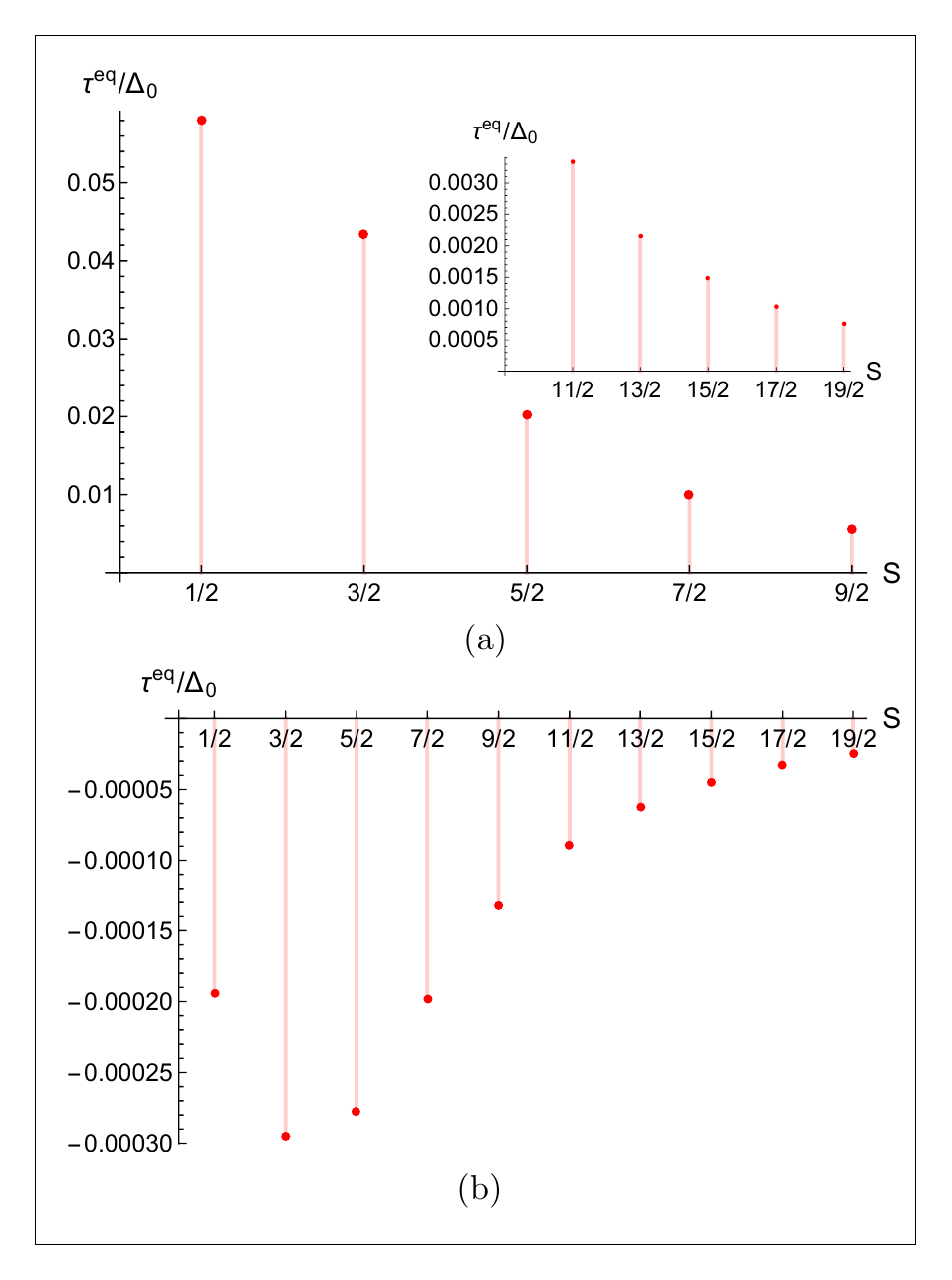}
\caption{\small \sl (a) Equilibrium quantum spin torque (EQST) vs spin ($S$) of spin flipper. (b) EQST vs spin flip probability ($f_{2}$) of spin flipper for $S=5/2$ and $m'=5/2 (f_{2}=0), m'=3/2$ and $m'=-5/2 (f_{2}=2.236), m'=1/2$ and  $m'=-3/2 (f_{2}=2.8284)$ and $m'= -1/2 (f_{2}=3)$. Parameters are $\Delta_{0}=1 meV$, $T/T_{c}=0.01$, $\varphi=\pi/2$, $J=1$, $m'=-1/2$ (for (a)), $Z=0$, $\theta\rightarrow0$, $h/E_{F}=0.5$. {EQST decreases with increase of spin $S$ of magnetic impurity.}}
\end{figure}
This finite $\tau^{eq}$ can be a check also on whether $SFFS$ junctions are clean or a contaminated with magnetic adatoms. In Fig.~6(b) we plot EQST as a function of exchange interaction $J$ from antiferromagnetic coupling ($J<0$) to ferromagnetic coupling ($J>0$) at phase difference $\varphi=\pi/2$. For $\theta\rightarrow 0$, ferromagnets have no role in flipping the electron's/hole's spin\cite{ping} and spin flip is only due to the spin flipper. We see that for ferromagnetic coupling there is no sign change of EQST with change in $J$. However, for antiferromagnetic coupling ($J < 0$) there is a sign change in $\tau^{eq}$ as $J$ changes from $J=0$ to $J=-2$, implying tunability of the sign of EQST via the exchange interaction of spin flipper.
\begin{figure}
\includegraphics[width=0.9\linewidth]{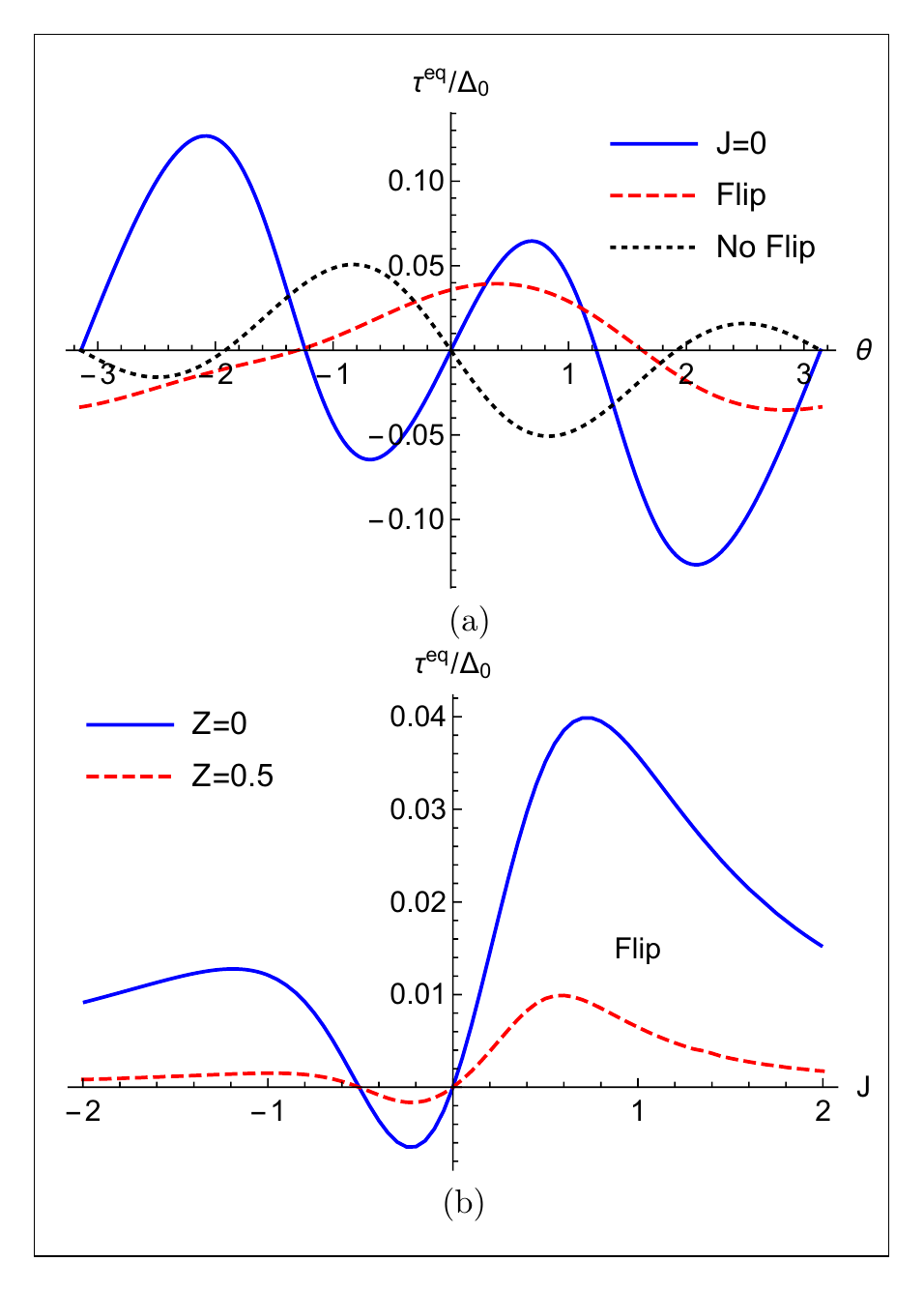}
\caption{\small \sl (a) EQST as a function of misorientation angle ($\theta$) for $\varphi=\pi/2$. (b) EQST as a function of exchange interaction ($J$) of spin flipper for $\varphi=\pi/2$ and $\theta\rightarrow 0$. Parameters are $\Delta_{0}=1 meV$,
$I_{0}=e\Delta_{0}/\hbar$, $T/T_{c}=0.01$, $Z=0$, $h/E_{F}=0.5$, spin flip case: $S=3/2, m'=-1/2$, no flip case: $S=3/2, m'=3/2$ and for (a) $J=1$. { In (a) EQST is zero for $J=0$ and no flip case ($f_{2}=0$), but finite for spin flip case ($f_{2}\neq0$). In (b) EQST changes sign with change in $J$ for antiferromagnetic coupling ($J<0$) and is also asymmetric with respect to $J$.}}
\end{figure}
Finally, in Fig.~7 we plot the EQST as a function of interface barrier strength ($Z$). We see that there is no sign change of EQST with increase of interface barrier strength $Z$. Further, the EQST is almost zero in the tunneling regime.\par{ The theoretically predicted numerical value of equilibrium spin transfer torque (ESTT) is $\sim10^{-2}$ meV in Ref.~\onlinecite{Wain}. On the other hand, in our work for the parameter values $Z=0$, $J=0.5$, $\varphi=\pi/2$, $S=5/2$ and $m'=-1/2$, the numerical value of equilibrium quantum spin torque (EQST) is $0.04$ meV. Thus, in our work the value of equilibrium quantum spin torque (EQST) is almost same with the value of equilibrium spin transfer torque as predicted in Ref.~\onlinecite{Wain}.}\par
{ Equilibrium spin current/torque in superconductor-ferromagnet-superconductor junctions with inhomogeneous magnetization is studied in Ref.~\onlinecite{alid}. They pointed out that there are discontinuous jumps in the equilibrium spin current or torque whenever the junction undergoes a $0-\pi$ transition. They find numerically that the spin current or torque is symmetric with respect to phase difference between two superconductors. They also show that for certain values of the thickness of ferromagnetic layer, a pure spin current can flow through the junction without any charge current. Similar to their work, we see in our work quantum spin torque is finite even when charge current vanishes. This finite quantum spin torque is antisymmetric with respect to phase difference between two superconductors in contrast to their work.}
\begin{figure}
\centering{\includegraphics[width=.7\linewidth]{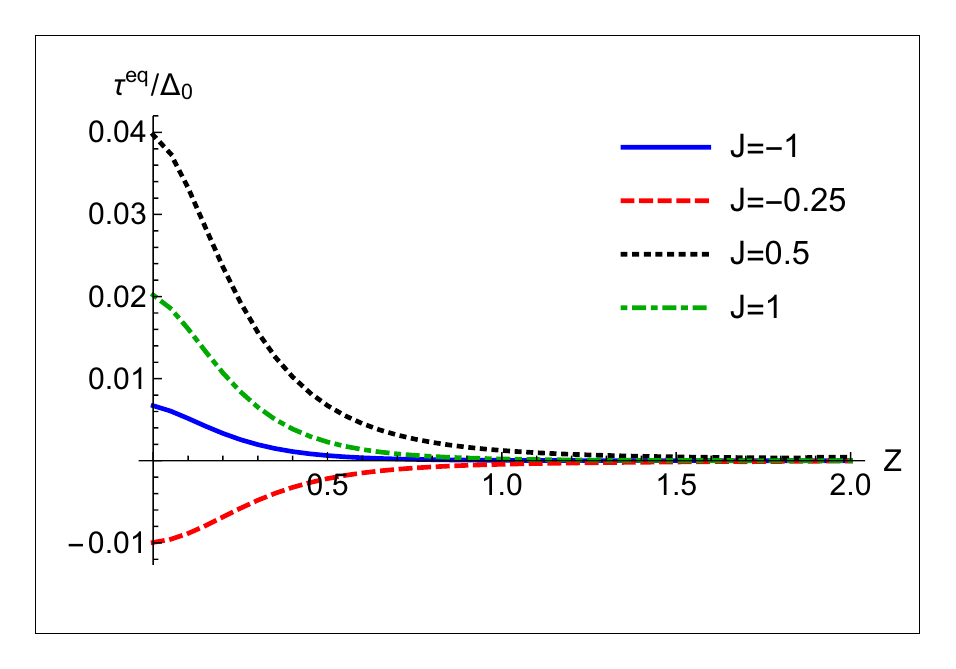}}
\caption{\small \sl EQST as a function of interface barrier strength ($Z$). Parameters are $\Delta_{0}=1 meV$, $I_{0}=e\Delta_{0}/\hbar$. {EQST decreases with increase of $Z$ and in the tunneling regime ($Z\rightarrow$large)  vanishes.}}
\end{figure}
\subsection{Physical picture: How does EQST arise?}
To understand the physical basis of the equilibrium quantum spin torque we go back to Fig.~2. When the Josephson super-current enters the first ferromagnetic layer { it} becomes spin-polarized in the direction of magnetization of the first ferromagnetic layer. This spin polarized {super-current} then interacts with the {magnetic impurity} through the exchange coupling and there is a finite probability for a mutual spin flip. One should note that this is a probability not a certainty, since the interaction of { spin polarized super-currents} is quantum in nature. Thus while before interaction the {super-current wave-function and {magnetic impurity} wave-function} are completely  independent after interaction  both are in a entangled and in a superposed state of: $\frac{m'}{2} |No-flip\rangle+\frac{f_{2}}{2}|Mutual-flip\rangle$, see paragraph below Eq.~\ref{eqq2}.

This finite probability of spin flip scattering implies the direction of the { super-currents} spin polarization is now too in a superposition of either polarized  in direction of the magnetization of ferromagnetic layers or not. Thus since the direction of the magnetization vector of both the ferromagnetic layers is same, say, this means the magnetization direction of second ferromagnetic layer will now differ from that of the { super-currents spin polarization state} which {is} in a superposition. Thus, when this {super-current} enters the second ferromagnetic layer, magnetic moment of the second ferromagnetic layer will exert a torque on that part of the { super-current wave-function} which is not in the same direction as the ferromagnet's, in order to rotate {its} spin state along the direction of magnetization, while leaving the non-spin flipped component of the {super-currents wave-function} as it is. From conservation of spin angular momentum, the spin flipped component of {super-currents wave-function} will also exert an equal and opposite torque on the magnetic moment of the second ferromagnetic layer. In this way, a torque arises although ferromagnet's are aligned. However, for no flip process, the wave-function is not in a superposition and in that case there is only a single no flip component. The {spin polarized state of the super-current} does not flip it's spin when interacting with the {magnetic impurity}. Thus in case of no-flip scattering the direction of the spin of { spin polarized super-current} and direction of magnetization of the ferromagnet's will remain the same. Thus equilibrium quantum spin torque vanishes in case of no-flip process but for spin-flip process it is finite. 
\section{Experimental realization and Conclusions}
The experimental detection of the novel phenomena pointed out in this work shouldn't be difficult. Superconductor-Ferromagnet-Ferromagnet-Superconductor (S-F-F-S) junctions have been fabricated experimentally for quite some time now\cite{Col}. Doping a magnetic ad-atom or magnetic impurity in S-F-F-S junctions with identical magnetization for ferromagnet's will experimentally implement our set up as shown in Fig.~2.
In conclusion, we have presented an exhaustive study of the nature of equilibrium spin torque in presence of a {magnetic impurity} of our hybrid system. We focus on the situation when the magnetic moments of the ferromagnetic layers are parallel. We identify spin flip scattering to be critical in inducing a torque in such a configuration. Further, we see that one can control the sign of this spin flip scattering induced Equilibrium quantum spin torque via exchange interaction of  as well as by the phase difference across the two superconductors. Tuning the sign of equilibrium quantum spin torque leads to control over the direction of magnetization of ferromagnet's. This has important implications in various spintronic devices as changing the direction of magnetization one can create faster magnetic random access memories\cite{Myer}.  
{\section{Appendix: Analytical expression for equilibrium quantum spin torque}}
{From Andreev bound states energies using Eq.~\ref{ff2} we can calculate the equilibrium spin torque ($\tau^{eq}$). For transparent regime ($Z=0$) we find-
\begin{equation}
\tau^{eq}\mid_{\theta\rightarrow 0}=\frac{\Delta_{0}^2}{2\pi}\int_{0}^{2\pi}\Big(\tanh\Big(\frac{\beta M_{1}}{2}\Big){M_{1}^{\prime}}+\tanh\Big(\frac{\beta M_{2}}{2}\Big){M_{2}^{\prime}}+\tanh\Big(\frac{\beta M_{3}}{2}\Big){M_{3}^{\prime}}+\tanh\Big(\frac{\beta M_{4}}{2}\Big){M_{4}^{\prime}}\Big)d(k_{F}a)
\label{eqn}
\end{equation}
\begin{eqnarray}
\mbox{ where } M_{1(2)}=&&\Delta_{0}\sqrt{D-\frac{1}{2}\sqrt{A+B}\pm\frac{1}{2}\sqrt{2A-B-\frac{2C}{\sqrt{A+B}}}},\nonumber\\ 
M_{1(2)}^{\prime}=&&-\frac{1}{2M_{1(2)}}\Big(-D^{\prime}-\frac{A^{\prime}+B^{\prime}}{4\sqrt{A+B}}\pm\frac{2A^{\prime}-B^{\prime}+\frac{C(A^{\prime}+B^{\prime})}{(A+B)^{3/2}}-\frac{2C^{\prime}}{(A+B)}}{4\sqrt{2A-B-\frac{2C}{\sqrt{A+B}}}}\Big),\nonumber\\
M_{3(4)}=&&\Delta_{0}\sqrt{D+\frac{1}{2}\sqrt{A+B}\pm\frac{1}{2}\sqrt{2A-B+\frac{2C}{\sqrt{A+B}}}}\nonumber\\ 
\mbox{ and }M_{3(4)}^{\prime}=&&-\frac{1}{2M_{3(4)}}\Big(-D^{\prime}+\frac{A^{\prime}+B^{\prime}}{4\sqrt{A+B}}\pm\frac{2A^{\prime}-B^{\prime}-\frac{C(A^{\prime}+B^{\prime})}{(A+B)^{3/2}}+\frac{2C^{\prime}}{(A+B)}}{4\sqrt{2A-B+\frac{2C}{\sqrt{A+B}}}}\Big),\nonumber
\end{eqnarray}
The explicit form of $A$, $B$, $C$, $D$, $A^{\prime}$, $B^{\prime}$, $C^{\prime}$, $D^{\prime}$ in Eq.~\ref{eqn} is
\begin{align}
\begin{split} 
A={}&4L_{1}^2-\frac{2}{3}L_{2},\\
B={}&\frac{2^{1/3}X_{1}}{3(X_{2}+\sqrt{X_{2}^2-4X_{1}^3})}+\frac{(X_{2}+\sqrt{X_{2}^2-4X_{1}^3})^{1/3}}{2^{1/3}3},\\
C={}&8L_{1}^{3}-2L_{1}L_{2}+L_{3},\\
D={}&L_{1},\\
A^{\prime}={}&-8L_{1}K_{1}-\frac{2}{3}K_{2},\\
B^{\prime}={}&\frac{2^{1/3}X_{1}^{\prime}}{3(X_{2}+\sqrt{X_{2}^2-4X_{1}^3})^{1/3}}-\frac{2^{1/3}X_{1}Y}{9(X_{2}+\sqrt{X_{2}^2-4X_{1}^3})^{4/3}},\\
C^{\prime}={}&-192L_{1}^{2}K_{1}+16L_{2}K_{1}-16L_{1}K_{2}+8K_{3},\\
D^{\prime}={}&K_{1},\\
\mbox{ where }\\
X_{1}={}&L_{2}^{2}-12L_{1}L_{3}-12L_{4},\\
X_{2}={}&2L_{2}^{3}-36L_{1}L_{2}L_{3}-432L_{1}^{2}L_{4}+27L_{3}^{2}+72L_{2}L_{4},\\
X_{1}^{\prime}={}&2L_{2}K_{2}+12L_{3}K_{1}-12L_{1}K_{3},\\
Y={}&Y^{\prime}+\frac{X_{2}Y^{\prime}-6X_{1}^{2}X_{1}^{\prime}}{\sqrt{X_{2}^2-4X_{1}^3}},\\\nonumber
\end{split}
\end{align}
\begin{align}
\begin{split}
Y^{\prime}={}&6L_{2}^2K_{2}-36L_{1}L_{3}K_{2}+36L_{2}L_{3}K_{1}+864L_{1}L_{4}K_{1}+54L_{3}K_{3}+72L_{4}K_{2},\\
L_{1}={}&P_{1}(S,m',f_{2},h,J,k_{F}a)+P_{2}(S,m',f_{2},h,J,k_{F}a)\cos(\varphi),\\
L_{2}={}&P_{3}(S,m',f_{2},h,J,k_{F}a)\cos(2\varphi)+P_{4}(S,m',f_{2},h,J,k_{F}a)\\
{}&\cos(\varphi)+P_{5}(S,m',f_{2},h,J,k_{F}a),\\
L_{3}={}&P_{6}(S,m',f_{2},h,J,k_{F}a)+P_{7}(S,m',f_{2},h,J,k_{F}a)\cos(\varphi)\\
{}&+P_{8}(S,m',f_{2},h,J,k_{F}a)\cos(2\varphi)+P_{9}(S,m',f_{2},h,J,k_{F}a)\\
{}&\cos(\varphi)\cos(2\varphi),\\
L_{4}={}&P_{10}(S,m',f_{2},h,J,k_{F}a)+P_{11}(S,m',f_{2},h,J,k_{F}a)\cos(\varphi)\\
{}&+P_{12}(S,m',f_{2},h,J,k_{F}a)\cos(2\varphi)+P_{13}(S,m',f_{2},h,J,k_{F}a)\\
{}&\cos(\varphi)\cos(2\varphi)+P_{14}(S,m',f_{2},h,J,k_{F}a)\cos(4\varphi),\\
K_{1}={}&f_{2}(P_{15}(S,m',f_{2},h,J,k_{F}a)\sin(\varphi)+P_{16}(S,m',f_{2},h,J,k_{F}a)),\\
K_{2}={}&f_{2}(P_{17}(S,m',f_{2},h,J,k_{F}a)+P_{18}(S,m',f_{2},h,J,k_{F}a)\cos(\varphi)\\
{}& +P_{19}(S,m',f_{2},h,J,k_{F}a)\sin(\varphi)+P_{20}(S,m',f_{2},h,J,k_{F}a)\sin(2\varphi)),\\
K_{3}={}&f_{2}(P_{21}(S,m',f_{2},h,J,k_{F}a)+P_{22}(S,m',f_{2},h,J,k_{F}a)\cos(\varphi)\\
{}&+P_{23}(S,m',f_{2},h,J,k_{F}a)\sin(\varphi)+P_{24}(S,m',f_{2},h,J,k_{F}a)\\
{}&\sin(2\varphi)+P_{25}(S,m',f_{2},h,J,k_{F}a)\sin(\varphi)\cos(2\varphi)).\nonumber
\end{split}
\end{align}
Here, $P_{i}$ ($i=1,2...25$) are functions of all parameters like exchange interaction ($J$), magnetization of the ferromagnets ($h$), spin ($S$) and magnetic moment ($m'$ ) of {magnetic impurity}, phase ($k_{F}a$) accumulated in ferromagnetic region and spin flip probability of {magnetic impurity} ($f_{2}$ ). Since these are large expressions we do not explicitly write them here. For no flip case- the spin flip probability of {magnetic impurity} is $f_{2}=0$. Thus, from above expressions: $K_{1}$, $K_{2}$, $K_{3}$ and also $A^{\prime}$, $B^{\prime}$, $C^{\prime}$ and $D^{\prime}$ vanish. Therefore, from Eq.~\ref{eqn}, $M_{1(2)}^{\prime}=0$ and $M_{3(4)}^{\prime}=0$, implying for no flip case equilibrium quantum spin torque vanishes ($\tau^{eq}\mid_{\theta\rightarrow0}=0$).}\\
\section{Ethics statement.} 
This work did not involve any collection of human or animal data.
\section{Data accessibility statement.}
This work does not have any experimental data.
\section{Competing interests statement}
We have no competing interests.
\section{Authors' contributions}
C.B. conceived the proposal, S.P. did the calculations on the advice of C.B., C.B. and S.P. analyzed the results and wrote the paper. Both authors reviewed the manuscript.
\section{Funding}
This work was supported by the grant ``Non-local correlations in nanoscale systems: Role of decoherence, interactions, disorder and pairing symmetry'' from SCIENCE \& ENGINEERING RESEARCH BOARD, New Delhi, Government of India, Grant No.  EMR/20l5/001836,
Principal Investigator: Dr. Colin Benjamin, National Institute of Science Education and Research, Bhubaneswar, India.

\end{document}